\documentstyle[aps,prl,preprint,tighten]{revtex}
\draft
\newcommand{\bc}{\begin{center}}
\newcommand{\ec}{\end{center}}
\newcommand{\nin}{\noindent}
\newcommand{\be}{\begin{equation}}
\newcommand{\ee}{\end{equation}}
\newcommand{\ba}{\begin{array}}
\newcommand{\ea}{\end{array}}

\newcommand{\dif}{{\rm d}}
\newcommand{\drho}{\rho^{\rm osc}}

\newcommand{\Gsc}{G_E^{\text sc}}
\newcommand{\cl}{\chi_{\scriptscriptstyle L}}
\newcommand{\kf}{k_{\scriptscriptstyle F}}

\newcommand{\br}{{\bf r}}
\newcommand{\A}{{\cal A}}
\newcommand{\C}{{\cal C}}

\begin{document}
\title{\bf $ $ \\
Orbital Magnetism in Ensembles of Ballistic Billiards}

\author{Denis Ullmo, Klaus Richter, and Rodolfo A. Jalabert \\
$ $ }

\address{Division de Physique Th\'eorique \footnote{Unit\'e de
Recherche des Universit\'es Paris XI et Paris VI associ\'ee au CNRS},
Institut de Physique Nucl\'eaire, F-91406 Orsay Cedex, France }

\date{Received 29 October 1993}
\maketitle
\begin{abstract}

We calculate the magnetic response of ensembles of small two-dimensional
structures at finite temperatures. Using semiclassical methods and numerical
calculation we demonstrate that only short classical trajectories are
relevant. The magnetic susceptibility is enhanced in regular systems, where
these trajectories appear in families. For ensembles
of squares we obtain a large paramagnetic susceptibility, in good agreement
with recent measurements in the ballistic regime.

\end{abstract}

\narrowtext

A free electron gas at temperature $T$ and magnetic field $H$ such that
$k_{B}T \gg \hbar w$ ($w=eH/mc$) exhibits a small diamagnetic response
\cite{Land}. This behavior persists when the electrons are placed in
periodic or weak-disorder potentials \cite{Peier}.
When the system is constrained to a finite volume the confining energy appears
as a relevant scale giving rise to finite-size corrections to the Landau
susceptibility. These corrections have been the object of several theoretical
studies in the last few years for the case of clean \cite{clean,Gefen} and
disordered \cite{disor} systems, and received renewed interest with
recent experiments of L\'evy {\it et al}. \cite{Levy}: Measurements on an
{\it ensemble} of $10^5$ microscopic, phase-coherent, ballistic \cite{phch}
squares lithographically defined on a high
mobility GaAs heterojunction yielded a large paramagnetic susceptibility at
zero field, decreasing on the scale of approximately one flux
quantum through each square. These experiments have been important in orienting
the theoretical studies towards the physically relevant questions associated
with the magnetic response of small systems. In particular, the role of finite
temperature and the necessity of distinguishing individual from ensemble
measurements appear as important ingredients that have been overlooked in
some of the theoretical literature.

In this letter we calculate the magnetic susceptibility  of noninteracting
electrons at finite temperatures in clean regular geometries (i.e. squares
and circles) for individual systems as well as for ensembles. We use
semiclassical approximation and classical perturbation theory since
the magnetic fields involved are not big enough to modify the classical
trajectories significantly. We explore the validity of our assumptions and
analytical results with numerical calculations. We compare the results
obtained for ensembles of regular structures with those of chaotic billiards,
finding important quantitative differences. We show that, within a
semiclassical approach, finite temperature induces a cut off on the classical
trajectories considered, and therefore clean systems can provide a good
description of the ballistic regime. This is the case for the experimental
conditions of Ref.~\cite{Levy} and therefore our model yields results in good
agreement with the measurements.

We consider an ensemble of isolated two-dimensional systems at temperature
$T$. For each member of the ensemble (with $N$ electrons and area $V$) the
magnetic susceptibility $\chi$ is given by the change of the free energy
$F(T,N,H)$ under the effect of a magnetic field,

\begin{equation}
\chi = - \frac{1}{V} \left(\frac{\partial^{2}F}{\partial H^{2}} \right)_{N,T}
 \ .
\label{eq:sus}
\end{equation}

\nin The necessity of using the canonical ensemble for isolated mesoscopic
systems,
and the physical differences with the grand-canonical ensemble (GCE, where the
system responds to the magnetic field with a fixed chemical potential $\mu$),
are some of the important concepts that recently emerged in the context of
persistent currents \cite{BM}. On the other hand, calculations in the GCE are
more easily performed due to the simple form of the thermodynamic potential

\begin{equation}
\Omega(T,\mu,H) = - \frac{1}{\beta} \int \ \dif E \ \rho(E) \
\ln{(1+\exp{[\beta(\mu\!-\!E)]})} \ ,
\label{eq:therpot}
\end{equation}

\nin in terms of the density of states $\rho(E) =
- (2/\pi) \ {\rm Im} \ g(E)$. The factor of 2 takes into account spin
degeneracy, $\beta=1/k_{B}T$, and $g(E)$ is the trace of the Green function
$G_{E}(\br',\br)$, i.e.

\be
g(E) = \int \dif \br \, G_E (\br,\br)  \; .
\label{trace}
\ee

Separating $\rho$ into a mean (Weyl) part, which is field independent, and
an oscillating part, $\rho(E) = \rho^{0}(E) + \drho(E)$,
we define a mean chemical potential $\mu^{0}$ from
$N = \int \dif E \rho(E) f(E\!-\!\mu) =
\int \dif E \rho^{0}(E)f(E\!-\!\mu^{0})$. \ ($f$ is the
Fermi-Dirac distribution function.) Considering that $\drho \ll \rho^{0}$,
it has been shown \cite{ensembl}

\begin{equation}
F(N) = F^{0} + \Delta F^{(1)} +\Delta F^{(2)} \ ,
\label{eq:fd}
\end{equation}

\nin where $F^{0} = \mu^{0}N + \Omega^{0}(\mu^{0})$ and $\Delta F^{(1)} =
\Omega^{\rm osc}(\mu^{0})$. We define $\Omega^{0}$ and $\Omega^{\rm osc}$
by using respectively $\rho^{0}$ and $\drho$ instead of $\rho$ in
Eq.~(\ref {eq:therpot}). The second-order term is \cite{ensembl}

\begin{equation}
\Delta F^{(2)} =  \frac{1}{2 \rho^{0}(\mu^{0})} \ \left[ \int \ \dif E
\ \drho(E) \ f(E\!-\!\mu^{0}) \right]^{2} \ .
\label{eq:df2}
\end{equation}

\nin $F^{0}$ is field independent and does not contribute to $\chi$.
$\Delta F^{(1)}$ gives the susceptibility in a GCE with chemical potential
$\mu^{0}$. In disordered systems it vanishes under impurity average, and we
will show that it is also the case within the averages of our semiclassical
model.

We will calculate $\drho$ from the semiclassical expansion
of the Green function. Except for a logarithmic
singularity when $\br'\! \rightarrow \! \br$, which yields the smooth part
$\rho^{0}$ of $\rho$, the semiclassical Green function has the generic
form \cite{gutz}

\be
\Gsc(\br',\br) = \sum_t D_t \exp{\left[\frac{i}{\hbar}
\left(S_t - \left(\eta_t-\frac{1}{2}\right)\frac{\pi}{2}\right)\right]} \ ,
\label{green}
\ee

\nin where the sum runs over all classical trajectories $t$ joining $\br$ to
$\br'$ {\em at energy $E$}. $S_t$ is the action integral along the trajectory.
For billiards without magnetic field we simply have $S_t\!=\!\hbar kL_{t}$,
$k=\sqrt{2mE}/\hbar$ and $L_t$ is the length of the trajectory.
The amplitude $D_t$ takes care of the classical probability conservation
and $\eta_t$ is the Maslov index.

Within our semiclassical approach, the free energy corrections
are given as sums over classical trajectories, each term being the
convolution in energy of the semiclassical contribution (oscillating as
$kL_t$) with the Fermi factor (smooth on the scale of $\beta$). It can be
shown \cite{Klaus} that the $T\!=\!0$ contribution to $\Delta F^{(1)}$ of a
trajectory is reduced by a temperature-dependent factor $R(T)=
(L_{t}/L_{c})\sinh^{-1}{(L_{t}/L_{c})}$, with
$L_{\rm c} = \hbar^{2} \kf \beta /(\pi m)$. A factor of $R^{2}(T)$ is
needed for $\Delta F^{(2)}$. For long trajectories and high temperatures,
$R(T)$ results in an exponential suppression and therefore the fluctuating
part of the free energy, and $\chi$, are dominated by trajectories with
$L_{t} \leq L_{\rm c}$, which will be the only ones considered in our
analysis. (We will not write $R(T)$ and $R^2(T)$ in the equations that follow.)

The standard route to obtain $\drho$ from $\Gsc$ is to evaluate the
integral of Eq.~(\ref{trace}) by stationary-phase approximation. This
selects the trajectories which are not only closed in configuration
space $(\br'\!=\!\br)$, but also closed in phase space ($\bf p'\! =\! \bf p$),
i.e.~periodic orbits.
When these latter are [well] isolated the Gutzwiller Trace Formula \cite{gutz}
is obtained. For integrable systems, periodic orbits come in continuous
families corresponding to the rational invariant tori (Berry-Tabor Trace
Formula \cite{bertab}). The difficulty in following this approach in our
case stems from the fact that in calculating $\chi$ for small fields, one is
actually looking at the effect of a small perturbation on rational tori.
The Poincar\'e-Birkhoff theorem states that, as soon as the field is turned on,
generically (the circular billiard being a
notable exception) all rational tori (i.e.~all families of periodic orbits)
are instantaneously broken, leaving only two (one stable, one unstable)
isolated periodic orbits. On the one hand, the physical effect which
generates $\chi$ is the breaking of the rational tori,
so that just ignoring this, i.e.~using the Berry-Tabor Formula,
is certainly inadequate. On the other hand, for $H\!\rightarrow \! 0$, the
remaining orbits are not sufficiently well isolated to apply the Gutzwiller
Trace Formula. Therefore, a uniform treatment of the perturbing field
is needed, where not only orbits that are closed in phase space are taken
into account, but also trajectories closed in configuration space which can be
traced to periodic orbits when $H\!\rightarrow \!0$.

In squares (of side $a$), due to the simplicity of the geometry, such a
uniform treatment is possible since we can perform the corresponding
integrals exactly. For $H \! = \! 0$, $\eta_t$ is twice the number
of reflections, and $D_t = \alpha / L_t^{1/2}$ with $\alpha =
- \pi (2m)^{3/4} /[(2\pi\hbar)^{3/2} E^{1/4}]$.
One way to obtain this result is to use the method of images and express
$G_E$ in terms of the free Green function $G_E^0$ as
	\be \label{image}
	G_E(\br',\br) = G_E^0(\br',\br)
	      + \sum_{\br'_i} \epsilon_i G_E^0(\br'_i,\br) \ ,
	\ee
where the $\br'_i$ are all the mirror images of $\br'$ by any combination
of symmetries across the sides of the square, and $\epsilon_i = \pm 1$
depending on the number of symmetries needed to map
$\br'$ on $\br'_i$. The long-range asymptotic behavior of the two-dimensional
free Green function $G_E^0(\br'_i,\br) \simeq \alpha
\exp{[i(k|\br'_i-\br|-\pi /4)]} / |\br'_i-\br|^{1/2}$
can be used for the images \cite{gutz2}.

For sufficiently weak magnetic fields,  one may keep in Eq.~(\ref{green}) the
zero-order approximation for $D_t$, and use the first-order correction
$\delta S$ to the action. For a closed orbit enclosing an algebraic area $\A$,
classical perturbation theory yields $\delta S = (e/c)H \A$ for low fields
and high energies, such that the cyclotron radius of the
electrons is much larger than the typical size of the structure.

We now specify the contribution $\rho_{11}$ [to $\drho$] of the family
of closed trajectories which, for $H\! \rightarrow \!0$, tends to the family of
shortest periodic orbits with non-zero enclosed area. We note it (1,1) since
the trajectories bounce once on each side of the square (upper inset,
Fig.~\ref{result1}). Their length is $L_{11}\! = \!2\sqrt{2} a$. This family
gives the main contribution to the experiment of Ref.~\cite{Levy} since
$L_{c}\! \approx \!2a$ at $T\!=\!40mK$. The contribution of other families is
obtained essentially in the same way. However, strong flux cancellation
occuring for other primitive orbits makes their contribution irrelevant in
the case of the square, even for very low temperatures \cite{Klaus,repet}.
Using as space coordinates $x_0$, which labels the
trajectory, $s$ the distance along the trajectory, and the index
$\epsilon = \pm 1$ specifying the sense of motion \cite{foot}, the
area is simply $\A_{\epsilon}(x_0) = \epsilon 2 x_0 (a-x_0)$.
Inserting $\A$ in Eq.~(\ref{trace}) we have
$\rho_{11}(H) = \rho_{11}({\scriptstyle H=0}) \C(H)$, where
$\rho_{11}({\scriptstyle H=0}) = - 8 a^2 \alpha \sin{(kL_{11}\!+\! \pi/4)} /
L_{11}^{1/2}$ is the unperturbed contribution and

	\be \label{C}
	\C(H) =
	  \int_0^{a} \dif x_0  \cos \left( \frac{2e}{\hbar c} H x_0 (a-x_0)
	  \right)
 	=
	  \frac{1}{\sqrt{2 \varphi}}
	  \left[ \cos(\pi \varphi) {\text C}(\sqrt{\pi \varphi}) +
	         \sin(\pi \varphi) {\text S}(\sqrt{\pi \varphi}) \right]
    	\; .
	\ee

\nin $\text C$ and $\text S$ are respectively the cosine and sine Fresnel
integrals, and $\varphi = \Phi/\Phi_0$ is the total flux $\Phi = Ha^2$ inside
the square measured in units of $\Phi_0 = hc/e$.
For $\varphi \geq 1$ the Fresnel integrals can be replaced by their
asymptotic value $1/2$, which amounts to evaluating $\C(\varphi)$ by
stationary phase, i.e. $\C^{\rm S} (\varphi) =
\cos(\pi \varphi \! + \! \pi/4) / \sqrt{4\varphi}$.
This expression however diverges for $H \! \rightarrow \! 0$, while $\C(0)=1$.

The contribution of the (1,1) family to $\Delta F^{(1)}$ yields, in
leading-order in $\kf a$

\be
\frac{\chi^{(1)}}{\cl}  = \frac{3}{(\sqrt{2}\pi)^{5/2}} \
(\kf a)^{3/2} \ \sin{\left(\kf L_{11}+
\frac{\pi}{4}\right)} \ \frac{\dif^2 \C}{\dif \varphi^2} \ .
\label{chi1}
\ee

\nin Therefore, the susceptibility of a given square can be paramagnetic or
diamagnetic
(Fig.~\ref{result1}) and its typical magnitude is much larger than $\cl$, with
$-\cl = -e^2/(12\pi mc^2)$ being the two-dimensional Landau susceptibility.
Clearly, $\chi^{(1)}$ vanishes under average if the dispersion of $\kf a$
across the ensemble is of the order of $2\pi$. The average
$\chi$ is then given by the contribution of the (1,1) family to
$\Delta F^{(2)}$

\be
\frac{\langle \chi \rangle}{\cl}  = - \frac{3}{(\sqrt{2}\pi)^3} \ \kf a
\ \frac{\dif^2 \C^2}{\dif \varphi^2} \ .
\label{chi}
\ee

\nin The average susceptibility (solid line, Fig.~\ref{result2}) is
paramagnetic at $H\!=\!0$ and for low fields it oscillates with an overall
decay of $1/\varphi$. The divergent susceptibility
obtained from $\C^{\rm S}$ (dotted line) provides a good description
of $\chi$ for $\varphi \geq 1$. For ensembles with a wide distribution
of lengths $a$ (in Ref.~\cite{Levy} the dispersion in size
across the array is estimated between 10 and 30\%) a second average in
$\dif^2 \C^2/\dif \varphi^2$ should be performed. (Since the scale of variation
of $\C$ with $a$ is much slower than that of $\sin{(\kf L_{11})}$ we can
effectively separate the two averages.) The low-field oscillations of
$\langle \chi \rangle$ are suppressed under the second average (done for
a gaussian distribution with a 30\% dispersion, dashed line),
while the zero-field behavior remains unchanged.

We checked the semiclassical results calculating the first 1500
eigenenergies of a square in a magnetic field by direct diagonalization.
At $T \! = \! 0$ the free energy reduces to the total
energy and $\chi$ is dominated by big paramagnetic singularities at the
level-crossings of states belonging to different symmetry classes and at
small avoided-crossings between states with the same symmetry \cite{clean}.
These peaks are compensated once the
next state is considered, and therefore disappear at finite temperature where
the occupation of nearly degenerate states becomes almost the same.
Temperature regularizes the $T \! = \! 0$ singular
behavior, and of course, describes the physical situation. We include
it by calculating the partition
function $Z = \exp{[-\beta F]}$ from a recursive algorithm
\cite{partfun,Klaus}.
The results for individual squares are in excellent agreement with
Eq.~(\ref{chi1}), the oscillations as a function of $\kf L_{11}$ (and
$\varphi$)
clearly shown in Fig.~\ref{result1}. The average values also agree with
our analytical findings (Fig.~\ref{result2}).

Ref.~\cite{Levy} yielded a paramagnetic susceptibility at $H\!=\!0$ with a
value of approximately 100 (whithin a factor of 4) in units of $\cl$. For the
two electron densities $n_{s}=10^{11}$ and $3 \! \times \! 10^{11} cm^{-2}$
of the experiment, the factor $4\sqrt{2}/(5\pi) \kf a$ from
Eq.~(\ref{chi}) gives respectively a susceptibility of 130 and 220, that when
temperature
is considered (through $R^2(T \!=\!40 mK)$) become 60 and 170, in
good agreement with the measurements. The field scale for the decrease of
$\langle \chi({\scriptstyle H=0}) \rangle$ is of the order of one flux quantum
through each square, in reasonable agreement with our theoretical findings.
The temperature scale
for the decrease of the susceptibility was identified as given by the inverse
time-of-flight $v_{\scriptstyle F}/a$, which is the same scale
$L_{t}/L_{c}$ that we find.

Squares constitute a generic example of an integrable system perturbed by a
magnetic field. It is interesting to compare our results with two extreme
cases: circles (which remain integrable under
the perturbation) and completely chaotic systems. Expressing the hamiltonian
of a circle (of radius $a$) in action-angle variables \cite{Keller},
$\drho$ can be written as a sum over families of periodic trajectories
\cite{bertab}. Within our
finite-temperature approach we restrict ourselves to the shortest ones, the
whispering-gallery trajectories who turn only once around the circle in
coming to the initial point after $M$ bounces. Their contribution to $\drho$ is

\be
\rho_{\rm wg}(H) = \sum_{M=3}^{\infty} \rho_{\scriptscriptstyle M}
({\scriptstyle H=0})
\cos{\left(\frac{eH}{\hbar c} A_{\scriptscriptstyle M} \right)} \ .
\label{dgwg}
\ee

\nin $\rho_{\scriptscriptstyle M}({\scriptstyle H=0})\!=\!
\sqrt{8}mL_{\scriptscriptstyle M}^{3/2}/(\sqrt{\pi}\hbar^2 k^{1/2}M^2)
\sin{(kL_{\scriptscriptstyle M}\!+\!\pi/4\!-\!3\pi M/2)}$ and
the length of the $M^{\rm th}$ trajectory is
$L_{\scriptscriptstyle M}=2Ma\sin{(\pi /M)}$, while
the enclosed area is $A_{\scriptscriptstyle M}=(Ma^2/2)\sin{(2\pi /M)}$.
$\chi^{(1)}$ oscillates as a
function of $\kf a$ with an amplitude
proportional to $(\kf a)^{3/2}$ (consistent with Ref.~\cite{Bog}) and
vanishes under ensemble average.
$\langle \chi({\scriptstyle H=0}) \rangle/\cl = (24/\pi) \kf a C$,
with $C = \sum_{M=3}^{\infty} \sin{(\pi/M)} \sin^{2}{(2\pi/M)}/M \approx 0.7.$
The sums over $M$ are rapidly convergent, indicating the dominance
of the first few periodic orbits.

Squares and circles give the same dependence on $\kf a$ for $\chi^{(1)}$ and
$\langle \chi \rangle$. This generic behavior for integrable systems can be
traced to the $k^{-1/2}$ dependence of the contribution to $\drho$ most
sensitive to the magnetic field. The numerical prefactors depend on the
specifics of the geometry. The main contribution to $\chi$ at $H\!=\!0$ comes
from interference between pairs of time-reversed trajectories. In the circle
all periodic orbits within a family have the same area, while for squares
the dominant family (1,1) also includes periodic orbits with small enclosed
area. This difference results in a larger value of
$\langle \chi({\scriptstyle H=0}) \rangle$ (by a factor of 10) for the circle
and the absence of the $1/\varphi$ damping of the low field oscillations.

For chaotic systems (of typical length $a$) with hyperbolic periodic
orbits the Gutzwiller Trace
Formula provides the appropriate path to calculate $\drho$. When only a
few short periodic orbits are important, $\chi$ can have any sign and
its magnitude is of the order of $\kf a
\cl$ \cite{Sha}. Extending this analysis to the
case of an ensemble of chaotic systems we obtain
$\langle \chi \rangle \propto \cl$. The individual $\chi$ are larger, by a
factor
$(\kf a)^{1/2}$ in regular geometries than in chaotic systems \cite{vOR}. For
$\langle \chi \rangle$ the difference is even larger (factor $\kf a$).
This is due to the large oscillations of $\rho$ in regular systems
induced by families of periodic trajectories.
The different magnetic response according to the geometry does not arise as a
long-time property (linear vs. exponential trajectory divergences) but as a
short-time property (family of trajectories vs. isolated trajectories). This
assures that small variations in the geometry of the clean systems
that we have considered will not be relevant.

We believe that measuring the susceptibility in different geometries will be of
high interest in order to understand the applicability of simple noninteracting
semiclassical models to actual microstructures.

We acknowledge helpful discussions with H.~Baranger, O.~Bohigas, Y.~Gefen,
M.~Gutzwiller, L.~L\'evy, N.~Pavloff, R.~Prange, B.~Shapiro, and
H.~Weidenm\"{u}ller. KR acknowledges financial support by the Humboldt
foundation.

\newpage

\begin{figure}
\caption{
Magnetic susceptibiltiy of a square as a function of $\protect \kf a$ from
numerical calculations at zero field and a temperature equal to 10
level-spacings. The number of
electrons is $N=(\protect \kf a)^2/(2 \pi)$. The
dashed line is the envelope of the oscillations (in $\protect \kf L_{11}$) of
our semiclassical approximation with the temperature correction factor $R(T)$.
The period $\pi/ \protect\sqrt{2}$ indicates the
dominance of the shortest periodic orbits enclosing non-zero area
with length $L_{11}=2\protect\sqrt{2} a$ (upper inset). Lower inset:
amplitude of the
oscillations (in $\protect \kf L_{11}$) of $\chi$ as a function of the flux
through the sample from Eq.~(\protect\ref{chi1}) (dashed) and numerics (solid).
}
\label{result1}
\end{figure}

\begin{figure}
\caption{
Average magnetic susceptibiltiy for an ensemble of squares from
Eq.~(\protect\ref{chi}) (solid) and from the stationary-phase integration
$\C^{\rm S}$ (dotted). Dashed: average over an ensemble with a large
dispersion of sizes (see text), Thick dashed: average from numerics.
Inset: average susceptibility as a function of $\protect \kf a$ for various
temperatures (4,6 and 10 level spacings) and a flux $\varphi=0.15$, from
Eq.~(\protect\ref{chi}) (dashed) and numerics (solid).
}
\label{result2}
\end{figure}


\begin{references}
\bibitem{phch} The size of the squares is $a=4.5$ $\mu m$. The
phase-coherence length is estimated between 15 and 40 $\mu m$ and the
elastic mean free path between 5 and 10 $\mu m$.
\bibitem{foot} Each point $\br'$ is counted four times,
corresponding to the four sheets of the invariant torus.
\end{references}
\end{document}